\documentclass[12pt]{article}
\usepackage{wallpaper}
\usepackage[margin=1in, paperwidth=8.5in, paperheight=11in]{geometry}
\usepackage{amsmath}
\usepackage{graphicx}%
\usepackage{amsfonts}%
\usepackage{amssymb}
\usepackage{color}
\usepackage{float,subfigure}
\usepackage[round]{natbib}
\usepackage{cite}
\usepackage{setspace}
\usepackage{comment}
\usepackage{url}
\usepackage{enumerate}
\usepackage{verbatim}
\usepackage{rotating}
\usepackage{microtype}

\newcommand{\bet}{ \mbox{\boldmath $ \eta $} }

\newcommand{\bbeta}{ \ensuremath{\boldsymbol{\beta}}}

\newcommand{\bphi}{ \mbox{\boldmath $\phi$}}

\newcommand{\btheta}{ \mbox{\boldmath $ \theta $} }

\newcommand{\Sig}{ \mbox{$\Sigma$} }
\newcommand{\sig}{ \ensuremath{\sigma}}

\newcommand{\bgamma}{ \mbox{\boldmath $\gamma$} }
\newcommand{\brho}{ \mbox{\boldmath $\rho$} }

\newcommand{\ind}{ \mbox{$\stackrel{\text{ind}}{\sim}$}}
\newcommand{\iid}{ \mbox{$\stackrel{\text{iid}}{\sim}$}}

\newcommand{\bzero}{\textbf{0}}

\newcommand{\bv}{ {\bf v} }

\newcommand{\bx}{ {\bf x} }

\newcommand{\bY}{ {\bf Y} }

\newcommand{\bZ}{ {\bf Z} }

\newcommand{\given}{\,\vert\,}

\newcommand{\tR}{\widetilde{R}}

\newcommand{\MCAR}{\mbox{$\text{MCAR}$}}
\newcommand{\MSTCAR}{\mbox{$\text{MSTCAR}$}}

\newcommand{\tbeta}{\mbox{$\text{Beta}$}}
\newcommand{\IW}{\mbox{$\text{InvWish}$}}
\newcommand{\IG}{\mbox{$\text{IG}$}}
\newcommand{\Wish}{\mbox{$\text{Wish}$}}
\newcommand{\Pois}{\mbox{$\text{Pois}$}}
\newcommand{\LSAA}{\mbox{$\text{SPY}$}}
\newcommand{\SPY}{\mbox{$\text{SPY}$}}
\newcommand{\Gam}{\mbox{$\text{Gamma}$}}

\begin{document}


\thispagestyle{empty}
\setcounter{page}{0}

\begin{center}
{\Large \textbf{Hierarchical Multivariate Space-Time Methods for Modeling Counts with an Application to Stroke Mortality Data}} 

\bigskip

\textbf{Harrison Quick$^{*1}$, Lance A.\ Waller$^{2}$, and Michele Casper$^1$}\\ 
$^{1}$ Division of Heart Disease and Stroke Prevention, Centers for Disease Control and Prevention, Atlanta, GA 30329\\
$^2$ Department of Biostatistics and Bioinformatics, Emory University, 
Atlanta, GA 30322\\
$^{*}$ \emph{email:} HQuick@cdc.gov

\end{center}

\textsc{Summary.}
{Geographic patterns in stroke mortality have been studied as far back as the 1960s, when a region of the southeastern United States became known as the ``stroke belt'' due to its unusually high rates of stroke mortality. While stroke mortality rates are known to
increase exponentially with age, an investigation of spatiotemporal trends by age group at the county-level is daunting due to the preponderance of small population sizes and/or few stroke events by age group.
Here, we harness the power of a complex, nonseparable multivariate space-time model which borrows strength across space, time, and age group to obtain reliable estimates of yearly county-level mortality rates from US counties between 1973 and 2013 for those aged 65+. Furthermore, we propose an alternative metric for measuring changes in event rates over time which accounts for the full trajectory of a county's event rates, as opposed to simply comparing the rates at the beginning and end of the study period. In our analysis of the stroke data, we identify differing spatiotemporal trends in mortality rates across age groups, shed light on the gains achieved in the Deep South, and provide evidence that a separable model is inappropriate for these data.
}

\textsc{Key words:}
{Age disparities in health}, Bayesian methods, Nonseparable models, Saved person-years, {Small area analysis} 

\newpage
\doublespacing
\section{Introduction}
Stroke (i.e., cerebrovascular disease) is the fourth leading cause of death in the general population and {third} --- behind heart disease and cancer --- among those aged 85 and older, with rates increasing exponentially with age \citep{deaths}. Geographic trends in stroke mortality have been studied as far back as the 1960s, when \citet{strokebelt} identified a region of the southeastern United States (US) stretching from Mississippi to North Carolina which had the highest rates of stroke mortality --- a region which would become known as the ``stroke belt.'' Later work by \citet{shiftingbelt} noticed an apparent shift in the stroke belt, observing that regions of the Mississippi River Valley appeared in the highest decile of mortality rates in the early 1990s where they had previously not. More recently, \citet{linda} have studied geographic trends in stroke hospitalizations from 1995 to 2006, noting that, among Medicare beneficiaries ages 65 and older, this shift in the stroke belt has persisted, stretching further into parts of Texas and Oklahoma.
In addition to changing geographic patterns, numerous studies have observed the overall declines in stroke mortality \citep[e.g.,][]{regards:declines,gillum:2011:stroke}. 

While many of these studies have age-adjusted their data, 
accounting for the variation in age distributions among counties and disparities in stroke mortality across age ranges, this 
precludes inference within individual age groups.
{On the other hand, this aggregation and data standardization step also helps mitigate the issue of small population sizes and low number of stroke deaths in many US counties, an issue that can lead to unreliable rate estimates
and is only exacerbated when the data are stratified by a factor such as age group.
In this work, however, our goal is to investigate spatiotemporal trends in stroke mortality by jointly modeling data from three age-based subpopulations, permitting more reliable inference at the county level for each age group while preserving the ability to compute age-adjusted rates.}
More specifically, we look to build a complex multivariate space-time model which
borrows strength across space, time, and age group.
{We also propose a new tool for measuring declines in mortality which accounts for temporal changes in mortality rates.}

To achieve this,
a natural starting point is the family of models based on the conditionally autoregressive (CAR) model proposed by \citet{besag}.  Since its extension to the fully Bayesian setting in \citet{bym}, CAR models have sparked a wealth of research
in the disease mapping context for both spatial \citep[e.g.,][]{besag:poisson,besag:higdon} and spatiotemporal applications. 
While these early examples were all based on the standard univariate CAR model, \citet{gelfand:mcar} 
developed methods for general multivariate CAR (MCAR) models, inspiring novel approaches for both multiple and spatiotemporal disease mapping {\citep[e.g.,][]{JBC,QBC,hcar,m-b:2013,b-r:2015}.} 
More recently, \citet{quick:waller} proposed a special case of the MCAR of \citet{gelfand:mcar} --- referred to as the multivariate space-time CAR ($\MSTCAR$) model --- for the purpose of analyzing county-level heart disease mortality rates in the US over time for various race/gender groups.
For a more complete coverage of the recent advances in spatial and space-time modeling, see \citet{BCG}. 

{
In any discussion of the spatiotemporal modeling literature, we would be remiss not to mention the subject of \emph{separability} --- i.e., models where the spatiotemporal covariance can be decomposed into the product of a purely spatial covariance and a purely temporal covariance.  While separable covariance structures offer computational benefits, \citet{stein2005} highlights some drawbacks of the separability assumption, and concerns over the utility of separable models have motivated the development of classes of nonseparable covariance functions in the univariate continuous space, continuous time setting \citep[e.g.,][]{cressie:huang, gneiting2002}.
In cases where both space and time are discrete, as encountered in this study,
\citet{knorr-held:2000} has discussed a variety of possible space-time interactions, but here again the focus has been on a single outcome, with extensions to the multivariate space-time setting being more recent developments.  In addition to the nonseparable $\MSTCAR$ proposed by \citet{quick:waller}, \citet{jon} implement a reduced rank multivariate spatiotemporal mixed effects model which is designed to analyze high dimensional data efficiently.  That said, both of these approaches restrict their attention to the case of Gaussian data. {Other methods \citep[e.g.,][]{JBC,m-b:2013} allow for varying spatial structures by utilizing \emph{proper} MCAR models --- while these approaches are feasible when the number of spatial regions in the spatial domain
is small, the large number of US counties 
prohibits the use of proper MCAR models.}
}

Here we extend the nonseparable $\MSTCAR$ model 
to the generalized linear model setting to analyze spatiotemporal trends in the dataset comprised of stroke mortality counts among those aged 65--74, 75--84, and 85+ described in Section~\ref{sec:data}. 
Specifically, due to the rarity of stroke deaths, 
the Gaussian assumption of \citet{quick:waller} may not be appropriate. 
As such, in Section~\ref{sec:methods} we detail our approach for embedding the $\MSTCAR$ model into a Poisson likelihood akin to \citet{bym}, {in addition to presenting the saved person-years (SPY) tool for measuring patterns in declines}.
We then analyze the stroke mortality data in Section~\ref{sec:anal}, where we {discover different spatiotemporal trends in mortality rates across age groups and observe evidence that a separable model would be inappropriate for these data.
Finally, we summarize our findings and offer some concluding remarks in Section~\ref{sec:disc}.}

\section{Data Description}\label{sec:data}
The study population for this analysis includes all US residents aged 65 or older. In order to assess differences across the high-risk age ranges, we separate our data into $N_g =3$ groups: those aged 65--74, those 75--84, and those 85$+$.
Annual counts of stroke-related deaths per county per age group were obtained from the National Vital Statistics System (NVSS) of the National Center for Health Statistics (NCHS).
Due to inconsistencies in the manner in which death records were recorded prior to 1973,
we restrict the analysis to data from 1973--2013 ($N_t = 41$ years) to ensure valid comparisons across time.
{Deaths from stroke were defined as those for which the underlying cause of death was cerebrovascular disease according to the 8th, 9th, and 10th revisions of the International Classification of Diseases 
(ICD; {ICD--8: 430--438}; ICD--9: 430--438; ICD-10: I60--69).
{Based on the {comparability ratios reported by \citet{icd8:icd9} and \citet{icd9:icd10}, which indicate a high degree of similarity between the three revisions of the ICD}, we assume that this definition is consistent over the 41 year study period.} 
The geographic unit used in this analysis was the county (or county equivalent).  Given changes in county definitions during the study period affecting ten counties (e.g., the merging/splitting of counties), a single set of $N_s$ = 3,099 regions (henceforth referred to as counties) from the contiguous lower 48 states (including the District of Columbia) was used for the entire study period. Annual population counts were based on the bridged-race intercensal estimates provided by {\citet{census}}. 
{While the number of individuals in each age bracket has risen considerably ({by 89\%, 101\%, and 267\%, respectively}), the number of stroke-related deaths for individuals 65--84 has decreased nearly 60\% since 1973. 
From a public health perspective, this reduction in stroke mortality is a great achievement.  From a statistical perspective, however,
this can lead to concerns over the reliability of county-level mortality rate estimates based on so few events. Figures illustrating the national population and death trends for each age group can be found in Web Appendix~B.
}

\section{Methods}\label{sec:methods}
\subsection{Statistical model}\label{sec:mstcar}
Letting $Y_i$ and $n_i$ denote the incidence of disease and the population at risk in county $i$, \citet{bym} proposed a model of the form
\begin{equation}
Y_{i} \sim \Pois\left(n_i \exp\left[\bx_i\bbeta + Z_i + \phi_i\right]\right), \text{for $i=1,\ldots,N_s$} \label{eq:bym}
\end{equation}
where $\bx_i$ denotes a $p$-vector of covariates with corresponding regression coefficients, $\bbeta$, $Z_{i}$ is a spatial random effect, and $\phi_i \ind N\left(0,\tau^2\right)$ is an exchangeable random effect.  In their work, \citet{bym} modeled $\bZ=\left(Z_1,\ldots,Z_{N_s}\right)'$ as arising from an intrinsic conditional autoregressive (CAR) model, which has the conditional distribution
\begin{equation}
Z_i\given \bZ_{(i)},\sig^2 \sim N\left(\sum_{j=1}^{N_s} w_{ij} Z_j \slash \sum_{j=1}^{N_s} w_{ij}, \sig^2\slash \sum_{j=1}^{N_s} w_{ij}\right) \label{eq:car}
\end{equation}
where $\bZ_{(i)}$ denotes the vector $\bZ$ with the $i$th element removed and $w_{ij}=1$ if $i$ and $j$ are neighbors {(denoted $i\sim j$)} and 0 otherwise. Recommendations for prior distributions for $\sig^2$ and $\tau^2$ are offered by \citet{bernardinelli} and \citet{waller:carlin}.

Extending~\eqref{eq:bym} and \eqref{eq:car} to a setting consisting of multiple spatial surfaces is straightforward.  Letting $Y_{ikt}$ denote the number of deaths in county $i$ during year $t$ for age group $k$, we model 
\begin{equation}
Y_{ikt} \sim \Pois\left(n_{ikt} \exp\left[\bx_{ikt}\bbeta_{kt} + Z_{ikt} + \phi_{ikt}\right]\right),  \label{eq:bym_k}
\end{equation}
for $i=1,\ldots,N_s$, $k=1,\ldots,N_g$, and $t=1,\ldots, N_t$ where $\phi_{ikt} \ind N\left(0,\tau_{k}^2\right)$. To account for the multivariate spatiotemporal association in the data,
we follow the MSTCAR model of \citet{quick:waller} --- itself a special case of the $\MCAR$ of \citet{gelfand:mcar} --- and let {$\bZ = \left(\bZ_{1\cdot\cdot}',\ldots,\bZ_{N_s\cdot\cdot}'\right)' \sim\MCAR\left(1,\Sig_{\eta}\right)$},
\begin{align*}
\pi\left(\bZ\given \Sig_{\eta}\right) &\propto \vert \Sig_{\eta}\vert^{-(N_s-1)\slash 2} \exp\left[-\frac{1}{2}\bZ'\left\{(D-W)\otimes\Sig_{\eta}^{-1}\right\}\bZ\right]\\
\text{and}\;\;\bZ_{i\cdot\cdot}\given \bZ_{(i)\cdot\cdot},\Sig_{\eta} &\sim N\left(\sum_{j\sim i} \bZ_{j\cdot\cdot} \slash m_i, \frac{1}{m_i}\Sig_{\eta}\right),
\end{align*}
where $\bZ_{i\cdot\cdot} = \left(\bZ_{i1\cdot},\ldots,\bZ_{iN_g\cdot}\right)'$, $\bZ_{ik\cdot} = \left(Z_{ik1},\ldots,Z_{ikN_t}\right)'$, $W$ is an adjacency matrix with elements $w_{ij}$, $D$ is an $N_s\times N_s$ diagonal matrix with elements $m_i = \sum_{j=1}^{N_s} w_{ij}$, $\Sig_{\eta}$ denotes the $N_tN_g \times N_tN_g$ covariance structure for our $N_t$ years and $N_g$ age groups and $\otimes$ denotes the Kronecker product.
The $\eta$ subscript is a reference to the construction of $\bZ$ from \citet{quick:waller} where the authors began by defining $\bv_{\iota\cdot t}\iid N(\bzero,G_t)$, where $\iota = 1,\ldots, N_s-1$, to be a collection of independent $N_g$-dimensional random variables with covariance $G_t$ for $\iota=1,\ldots,(N_s-1)$ and $t=1,\ldots, N_t$.  Using $R_k = R\left(\rho_k\right)$ to denote an age group-specific temporal correlation matrix based on an autoregressive order 1 (denoted AR(1)) model and letting $\tR_k$ to be the Cholesky decomposition of $R_k$ such that $\tR_k \tR_k' = R_k$, the authors define $\bet_{\iota k\cdot} = \tR_k \bv_{\iota k\cdot}$ where $\bv_{\iota k\cdot} = \left(v_{\iota k1},\ldots,v_{\iota kN_t}\right)'$.  We then find that $\bet_{\iota\cdot\cdot} \sim N\left(\bzero,\Sig_{\eta}\right)$, where
\begin{align}\label{eq:Sig_eta}
\Sig_{\eta} =
\begin{bmatrix}
\tR_{1,1}^* & \bzero & \bzero\\
\vdots & \ddots & \bzero \\
\tR_{N_t,1}^* & \cdots & \tR_{N_t,N_t}^*
\end{bmatrix}
\begin{bmatrix}
G_1 & \bzero & \bzero\\
\bzero & \ddots & \bzero \\
\bzero & \bzero & G_{N_t}
\end{bmatrix}
\begin{bmatrix}
\tR_{1,1}^* & \cdots & \tR_{N_t,1}^*\\
\bzero & \ddots & \vdots \\
\bzero & \bzero & \tR_{N_t,N_t}^*
\end{bmatrix}
\end{align}
and $\tR_{t,t'}^*$ denotes the $N_g\times N_g$ diagonal matrix with elements $\left\{\tR_k\right\}_{t,t'}$ for $k=1,\ldots,N_g$; $\bZ$ is then constructed from the $\bet_{\iota\cdot\cdot}$ using the eigenvalues and eigenvectors of the matrix $D-W$ \citep[see][]{rue:held}. This structure is then denoted $\bZ \sim \MSTCAR\left(G_{1},\ldots,G_{N_t},\brho\right)$.


\subsection{Hierarchical model and computational details}
While a Poisson model like~\eqref{eq:bym_k} is a straightforward extension of~\eqref{eq:bym}, such models can also pose computational challenges, particularly for large dimensions.  For instance, the full conditional of $\bZ_{i\cdot\cdot}$, given by
\begin{align}
\pi\left(\bZ_{i\cdot\cdot}\given \bY,\bZ_{(i)\cdot\cdot},\bbeta,\bphi,\Sig_{\eta}\right) \propto& \prod_{k=1}^{N_g} \prod_{t=1}^{N_t} \Pois\left(Y_{ikt}\given n_{ikt} \exp\left[\bx_{ikt}\bbeta_{kt} + Z_{ikt} + \phi_{ikt}\right]\right) \notag\\
&\times \pi\left(\bZ_{i\cdot\cdot}\given \bZ_{(i)\cdot\cdot},\Sig_{\eta}\right)
\end{align}
is \emph{not} a known distribution.  That is, if we use a Markov chain Monte Carlo (MCMC) algorithm to estimate the posterior distribution of our model parameters, this model may require the use of large multivariate Metropolis updates within our Gibbs sampler. \citet{besag:poisson} and \citet{knorr-held:rue} suggest a reparameterization of~\eqref{eq:bym_k} which involves integrating $\phi_{ikt}$ out of the model, yielding a Gaussian full conditional for $\bZ_{i\cdot\cdot}$ and requiring Metropolis updates for $\theta_{ikt} = \bx_{ikt}\bbeta_{kt} + Z_{ikt} + \phi_{ikt}$.  Fortunately, the $\theta_{ikt}$ are independent of one another given $\bY$ and the other model parameters, so these Metropolis updates can be conducted independently and in parallel.

We complete our hierarchical model by specifying the following prior distributions for our other model parameters:
a vague 
prior for $\bbeta$, a weakly informative inverse Gamma prior for each $\tau_k^2$, a beta prior for each $\rho_k$, and an inverse Wishart prior for each $G_t$ with hyperparameter $G$, itself modeled using a Wishart prior.
While this 
structure on the covariance matrices
is likely unnecessary given the number of spatial regions in the data \citep[see the discussion of prior sensitivity in spatial models by][]{bernardinelli}, this 
comes at little-to-no computational cost (see {Web Appendix~A.6}) and offers a convenient means for specifying proper priors. 
Putting these pieces together, our full hierarchical model is as follows:
\begin{align}\label{eq:hier}
\pi\left(\bbeta,\bZ,G,G_1,\ldots,G_t,\brho,\tau_k^2,\btheta\given\bY\right) \propto&
\prod_{i,k,t} \Pois\left(Y_{ikt}\given n_{ikt}\exp\left[\theta_{ikt}\right]\right)
\times N\left(\btheta\given X\bbeta+\bZ,\Sig_{\theta}\right)\notag\\
&\times \MSTCAR\left(\bZ\given G_{1},\ldots,G_{N_t},\brho\right) \notag\\
&\times \prod_{t=1}^{N_t} \IW\left(G_t\given G,\nu\right) \times \Wish\left(G\given G_0,\nu_0\right)\notag\\
&\times \prod_{k=1}^{N_g} \left[\tbeta\left(\rho_k\given a_{\rho},b_{\rho}\right) \times \IG\left(\tau_k^2\given a_{\tau},b_{\tau}\right)\right] ,
\end{align}
where $\Sig_{\theta}$ is a diagonal matrix of size $N_sN_gN_t$ with elements $\tau_{k}^2$ and $X$ is the $(N_sN_gN_t\times p)$ matrix of covariates.

{While full details for implementing this model in an MCMC framework are provided in {Web Appendix~A}, we would be remiss to not discuss the computational burden associated with fitting a nonseparable model as opposed to a {separable} model; i.e., letting $\rho_k = \rho$ for $k=1,\ldots,N_g$ and $G_t = G$ for $t=1,\ldots,N_t$ corresponds to fitting a separable multivariate space-time model with $\Sig_{\eta} = R\left(\rho\right) \otimes G$.  First note that by using an AR(1) model for time, we can compute the $\tR_{t,t'}^*$ elements of $\Sig_{\eta}$ in closed-form, reducing the burden of computing $\Sig_{\eta}^{-1}$ from an $N_tN_g \times N_tN_g$ matrix inversion to a series of $N_g\times N_g$ matrix inversions. Furthermore, while the nonseparable $\MSTCAR$ model contains more parameters than its separable counterpart, the additional computational burden associated with its implementation in an MCMC framework is negligible.  Specifically, the computations necessary to construct the full conditional distributions for each $G_t$ are simply a partitioning of those necessary for constructing the full conditional distribution of $G$ in a separable model.}

When implementing this model in an MCMC framework, we have found that proper specification of the initial values can be crucial to facilitating convergence in a timely fashion. In particular, the parameters which require Metropolis updates --- $\rho_k$ and $\theta_{ikt}$ --- should be treated with care. For instance, we recommend initializing $\rho_k$ to be large (say, 0.90) if a high degree of temporal correlation is expected.
More importantly, 
we recommend initializing
\begin{align}
\theta_{ikt}^{(0)} =
\begin{cases}
\log \left(Y_{ikt}\slash n_{ikt}\right), &\text{if $Y_{ikt} > \epsilon$}\\
\log \left(\sum_{i} Y_{ikt} \slash \sum_{i} n_{ikt}\right), &\text{if $Y_{ikt}\le \epsilon$}
\end{cases},
\end{align}
where $\epsilon\ge 0$ is some small nonnegative integer, as this will allow the model to learn about parameters such as $\bbeta$ and $\bZ$ early on in the process; in practice, letting $\epsilon=0$ has been sufficient.  When $\theta_{ikt}$ is poorly initialized, however, the MCMC algorithm may take a large number of iterations to recover,
resulting in a chain which is slow to converge.  

{
\subsection{Assessment of reliability}\label{sec:reliable}
The primary motivation for this work is to achieve more reliable age-specific mortality rate estimates for these data.  In order to assess the reliability of our estimates, we will begin by generating \emph{synthetic} death counts from the posterior predictive distribution for $Y_{ikt}^{*}$,
\begin{align*}
Y_{ikt}^{*(\ell)}\given \theta_{ikt}^{(\ell)} \sim \Pois\left(n_{ikt}\exp\left[\theta_{ikt}^{(\ell)}\right]\right), \text{for $i=1,\ldots,N_s$, $j=1,\ldots,N_t$, $k=1,\ldots,N_g$},
\end{align*}
and for $\ell=1,\ldots,L$, where $\theta_{ikt}^{(\ell)}$ denotes the $\ell$th (post-burn-in) sample from the posterior distribution of $\theta_{ikt}$. We can then compute the 95\% CI for $Y_{ikt}^*$ from these posterior predictions and determine the proportion of each county's $N_tN_g$ 95\% CIs that contain the observed $Y_{ikt}$ to estimate the coverage probability. As a baseline for comparison purposes, we will compare our results to those generated from an empirical Bayesian Poisson-gamma model of the form
\begin{align}
\bgamma\given \bY,a_{kt},b_{kt} \propto \prod_{ikt} \Pois\left(Y_{ikt}\given n_{ikt}\gamma_{ikt}\right) \times \Gam\left(\gamma_{ikt} \given a_{kt},b_{kt}\right)\label{eq:poisgam}
\end{align}
where $a_{kt} = \sum_{i} Y_{ikt} \slash \sum_i n_{ikt} \times b_{kt}$ and $b_{kt} = 1000$, indicating a prior distribution equivalent to 1,000 additional persons with an event rate equal to the national average. We can then generate synthetic death counts from the resulting posterior predictive distribution, $Y_{ikt}^{\dagger (\ell)}\given \gamma_{ikt}^{(\ell)} \sim \Pois\left(n_{ikt}\gamma_{ikt}^{(\ell)}\right)$ as before.  The method which yields coverage probabilities near 0.95 more consistently will be deemed more reliable.
}

\subsection{Tools for measuring temporal changes in mortality}
When studying temporarily-varying mortality rates, it is often of interest to measure the decline from the beginning of the study period to the end.  Letting $\lambda_{ikt} = \exp\left[\theta_{ikt}\right]$ denote the county-specific mortality rate for group $k$ at time $t$ and letting $\Delta_{ik}(t,t') = \left(\lambda_{ikt} - \lambda_{ikt'}\right)\slash\lambda_{ikt}$ denote the percent change from time $t$ to $t'> t$ for group $k$ in county $i$, an obvious choice would be to compute $\Delta_{ik}(1,N_t)$ for each county and each group.  Similarly, one could define $\Delta_{i}(t,t') = \left(\sum_{k} n_{ikt}\lambda_{ikt} - \sum_{k} n_{ikt'} \lambda_{ikt'}\right)\slash \sum_{k} n_{ikt}\lambda_{ikt}$ and compute $\Delta_{i}(1,N_t)$ as an estimate of the county's percent decline.

The drawback of these quantities is that they only account for the rates at the beginning and end of the study period, ignoring the intervening periods.  Here, we propose a measure we refer to as the
``saved person-years'' --- or SPY ---
which can be computed as
\begin{align}\label{eq:lsaa}
\SPY_i &= \frac{1}{N_t-1} \sum_{t=2}^{N_t} \frac{\sum_k n_{ikt} \left(\lambda_{ik1} \left[1-\Delta_{\cdot k}(1,t)\right] - \lambda_{ikt}\right)}{\sum_{k} n_{ikt}} \times \text{100,000},
\end{align}
where $\Delta_{\cdot k}(1,t) = \left(\lambda_{\cdot k1} - \lambda_{\cdot k N_t}\right)\slash \lambda_{\cdot k 1}$ denotes the nationwide average decline from time 1 to time $t$ with $\lambda_{\cdot kt} = \sum_i n_{ikt} \lambda_{ikt} \slash \sum_i n_{ikt}$.
Note that $\SPY_i$ is essentially a measure of the deviation between the expected rate for county $i$ at time $t$ if the county had declined at the rate of the national average --- i.e., $\lambda_{ik1} \left[1-\Delta_{\cdot k}(1,t)\right]$ ---
and the rate we estimate from the model. While this quantity should not replace the crude measure of the percent decline, it is a simple and easily interpretable investigative tool that
tells a more thorough story
of a region's trajectory over the study period, as we will demonstrate in our analysis. 

In practice, we estimate $\theta_{ikt}$ by obtaining samples from its posterior distribution --- say, $\theta_{ikt}^{(\ell)}$ for iteration $\ell$ of our MCMC algorithm.  As such, both $\Delta_{i}(t,t')$ and $\LSAA_i$ can be computed for each iteration of our MCMC algorithm, resulting in posterior distributions for each of these quantities.  From this point forward, however, we restrict our attention to the posterior medians of $\Delta_{i}(t,t')$ and $\LSAA_i$ unless otherwise stated.


\section{Analysis of the Stroke Mortality Data}\label{sec:anal}
In the absence of covariate information, we fit the hierarchical model in~(\ref{eq:hier}) to the stroke mortality data described in Section~\ref{sec:data} using 
an intercept term for each combination of year and age group \citep[as required when using an improper CAR model, per][]{besag95}, forcing the random effects to account for a substantial amount of the spatio-temporal variability in the data.
We place an informative beta prior on each $\rho_k$ to encourage higher temporal correlations in the model,
and we use a vague inverse Wishart prior for each of the $G_t$.
When running the MCMC algorithm,
we thinned our posterior samples for $\theta_{ikt}$ by removing 9 out of 10 samples --- while this is not theoretically necessary, it {reduced the burden of storing excess samples for our nearly 400,000 random effects}. Estimates provided are based on posterior medians, and 95\% credible intervals (95\% CI) were obtained by taking the 2.5- and 97.5-percentiles from the thinned post-burn-in samples. Additional figures, including animations displaying temporal evolutions in the geographic trends, can be found in {Web Appendix~B}.

Before delving into the epidemiologic findings, we evaluate some of the numerous variance parameters permitted by the use of a nonseparable model. While the $\MSTCAR$ model in Section~\ref{sec:mstcar} is derived using temporally-varying covariance matrices, $G_t$, these parameters are not necessarily of direct interest as they are the variance parameters for $\bv_{\ell\cdot t}$, and thus they are \emph{not} directly interpretable with respect to the mortality rates. Instead, we need to use the posterior samples of $G_t$ and $\rho_k$ to construct $\Sig_{\eta}$ from~\eqref{eq:Sig_eta}.  These values coincide with the conditional covariance matrix of $\bZ_{i\cdot\cdot}$ (when scaled by the number of neighbors, $m_i$), and thus \emph{are} {interpretable with respect to the log mortality rates}.
{Furthermore, patterns in these parameters can be easily interpreted, as well.
For instance, Figure~\ref{fig:var} displays the posterior estimates for the diagonal elements of $\Sig_{\eta}$ corresponding to each age group.
Here, the declines in Figures~\ref{fig:var2} and~\ref{fig:var3} suggest a higher degree of spatial smoothing in later years than at the beginning of the study period, as the corresponding $Z_{ikt}$ become less free to deviate from their neighbors over time. As we will see, this may be in part due to declines in the mortality rates, themselves, as lower rates may also imply a smaller range of rates.}

\begin{figure}[t]
    \begin{center}
        \subfigure[Ages 65--74]{\includegraphics[width=.32\textwidth]{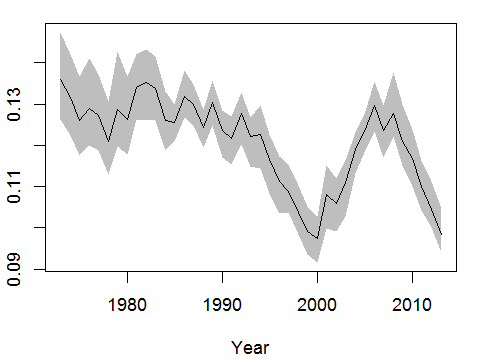}\label{fig:var1}}
        \subfigure[Ages 75--84]{\includegraphics[width=.32\textwidth]{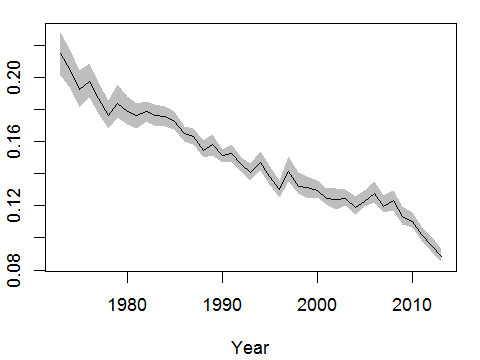}\label{fig:var2}}
        \subfigure[Ages 85+]{\includegraphics[width=.32\textwidth]{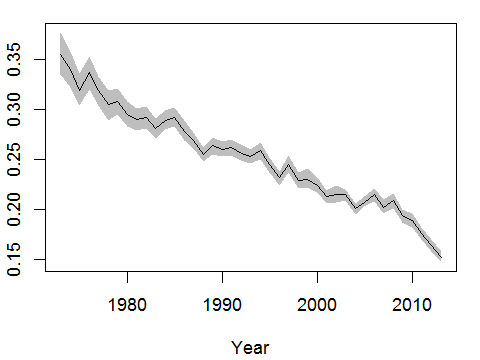}\label{fig:var3}}
    \end{center}
    \caption{Temporal evolution of the diagonal elements of $\Sig_{\eta}$.  While the scale of these parameters is not directly interpretable, declines in these variance parameters suggest an increase in spatial smoothing over time (particularly for those 75+).} 
    \label{fig:var}
\end{figure}

{
To assess the reliability of our estimates, we follow the approach set forth in Section~\ref{sec:reliable}. We begin by generating $L=1000$ replicates for each $Y_{ikt}$ from the posterior predictive distribution corresponding to both the $\MSTCAR$ model fit in~\eqref{eq:hier}, as well as the empirical Bayesian Poisson-gamma model from~\eqref{eq:poisgam}. After computing the 95\% CI of the replicates from both models and comparing these intervals to the observed $Y_{ikt}$, we find that the mean county-specific coverage from the $\MSTCAR$ is 97.7\%, whereas the Poisson-gamma yields a coverage of 99.6\%.  What is immediately clear from this result is that while the intervals from the $\MSTCAR$ are consistently more narrow than those from the Poisson-gamma model, this increase in precision due to the $\MSTCAR$ model is valid.  {Figures related to this reliability assessment are included in Web Appendix~B}.
}

Having demonstrated the necessity and utility of the $\MSTCAR$ model for these data, we now shift our attention to the rate estimates themselves.
For the sake of illustration of the temporal trends and the $\SPY$ tool, we highlight two counties: one from the heart of the stroke belt (Jefferson County, AL), and one from the opposite side of the country (King County, WA).
Figure~\ref{fig:gt} displays time trends of the estimated rates for these two counties
for each of the age groups, along with the national averages.  Here, we observe that while these counties exhibit similar trends for the 65--74 age group --- where King County consistently outperforms both Jefferson County and the nation as a whole --- the trends for the remaining age groups are quite different.  This is particularly true for the eldest age group in our study, as Jefferson County experienced such a sharp decline from 1973 to 1990 so as to not only pass King County, but also the national average.  This period of consistent declines was followed by stagnant rates through the early 1990s and an \emph{increase} in rates until the early 2000s, leading to the county ending the study period among the worst in the nation {(albeit with much lower rates than in the 1970s)}.

\begin{figure}[t]
    \begin{center}
        \subfigure[Ages 65--74]{\includegraphics[width=.32\textwidth]{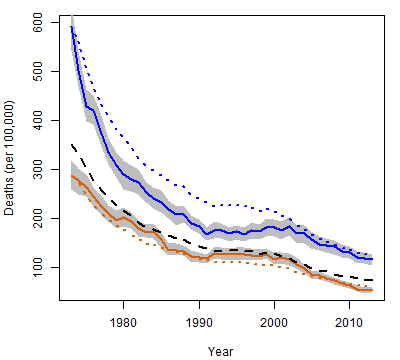}\label{fig:gt1}}
        \subfigure[Ages 75--84]{\includegraphics[width=.32\textwidth]{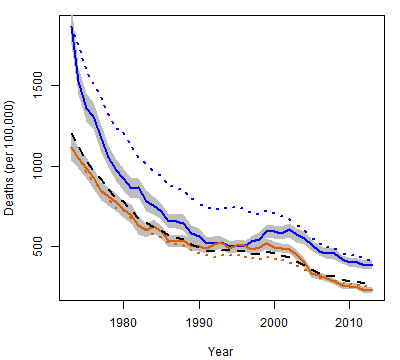}\label{fig:gt2}}
        \subfigure[Ages 85+]{\includegraphics[width=.32\textwidth]{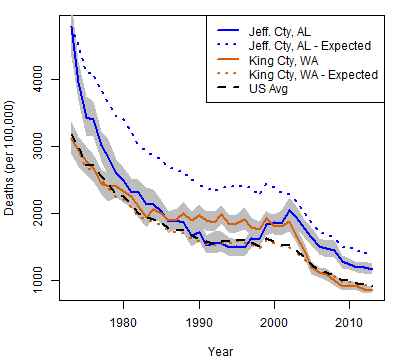}\label{fig:gt3}}
    \end{center}
    \caption{Comparison of the rate trends over time for two selected counties (solid lines), along with the national average (dashed line).  Gray bands denote 95\% CI. Also displayed for each county is the expected rate had the county declined at the same rate as the national average (dotted lines). 
    }
    \label{fig:gt}
\end{figure}

One aspect highlighted by these figures, however, is that computing a simple percent decline does not tell the whole story.  For instance, while Jefferson County experienced some degree of increasing rates in each of the three age groups during the late 1990s, simply measuring the age-standardized decline from 1973 to 2013 would overlook the strides the county took during the first half of the study period, when Jefferson County declined at a rate much faster than the national average.  The same cannot be said, however, for King County,
which underperformed during the first 30 years of the study, despite experiencing an overall decline that outperformed the national average.
This discrepancy can be observed by investigating the $\LSAA_i$ tool for each county.
Here, we find that Jefferson County saved 136.6 (98.8, 174.9) person-years per 100,000, while the $\SPY_i$ for King County was -34.0 (-61.7, -5.3) person-years per 100,000, reinforcing our claim that 
King County's declines lagged behind the national average.
A map of the $\LSAA_i$ values for all 3,099 counties can be found in Figure~\ref{fig:lsaa}, where we find that parts of the Deep South outperformed much of the nation, and a comparison
to the percent decline can be found in Figure~B.4 of the Web Appendix.}

\begin{figure}[t]
    \centering
        \includegraphics[width=.5\textwidth]{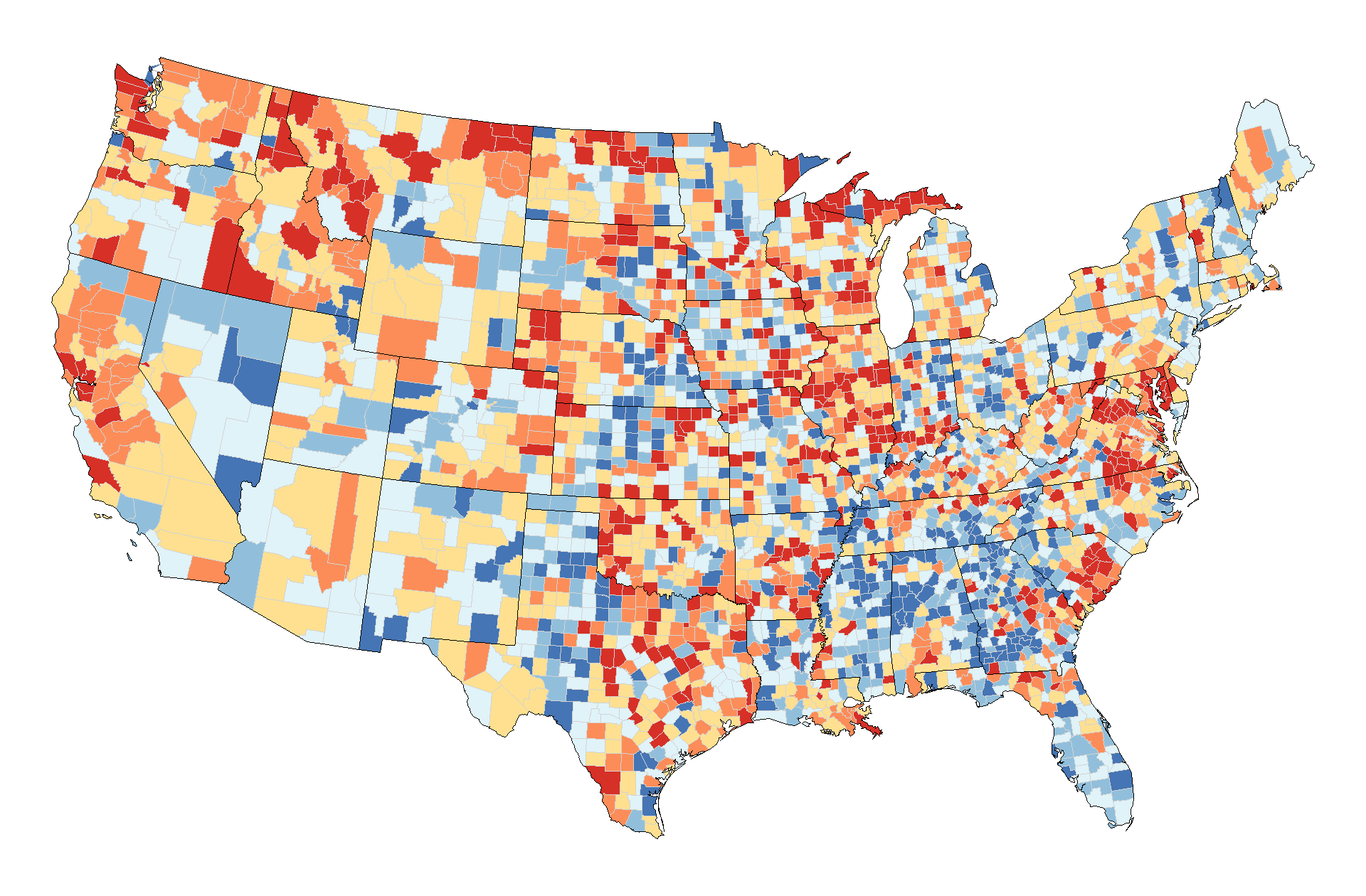}
        \includegraphics[width=.10\textwidth]{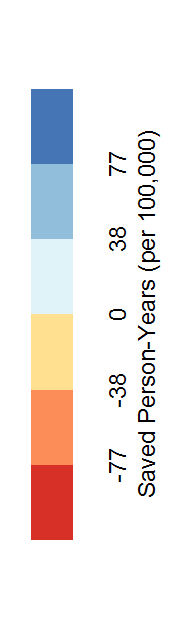}
    \caption{Map of the saved person-years (SPY)
    measure from~\eqref{eq:lsaa}, which measures the average difference between the model-estimated mortality rate of a county (per 100,000) and the expected mortality rate assuming a rate of decline equivalent to the national average.} 
    \label{fig:lsaa}
\end{figure}

\begin{sidewaysfigure}[t]
    \begin{center}
        \subfigure[Ages 65--74: 1973]{\includegraphics[width=.32\textwidth]{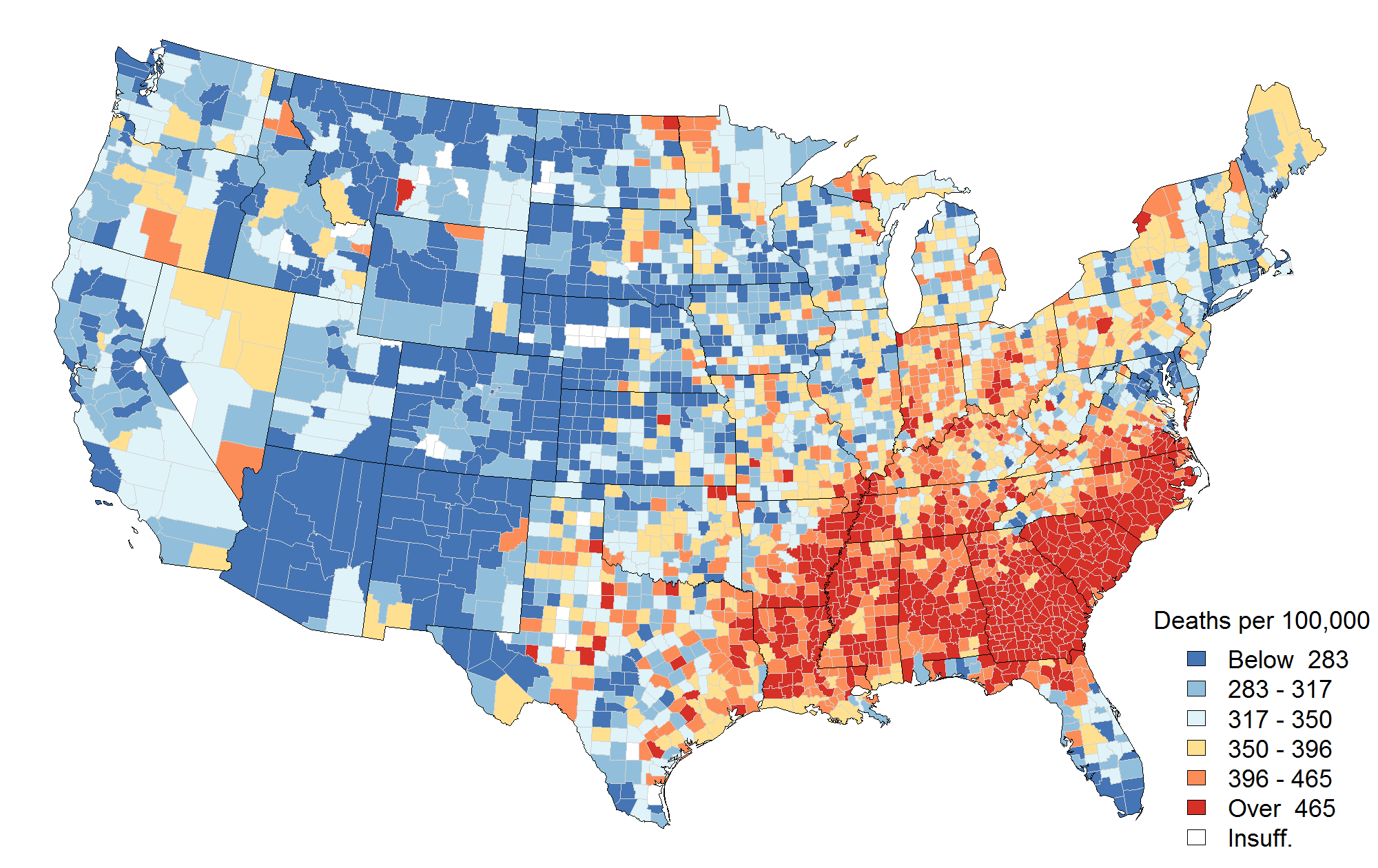}\label{fig:6a}}
        \subfigure[Ages 75--84: 1973]{\includegraphics[width=.32\textwidth]{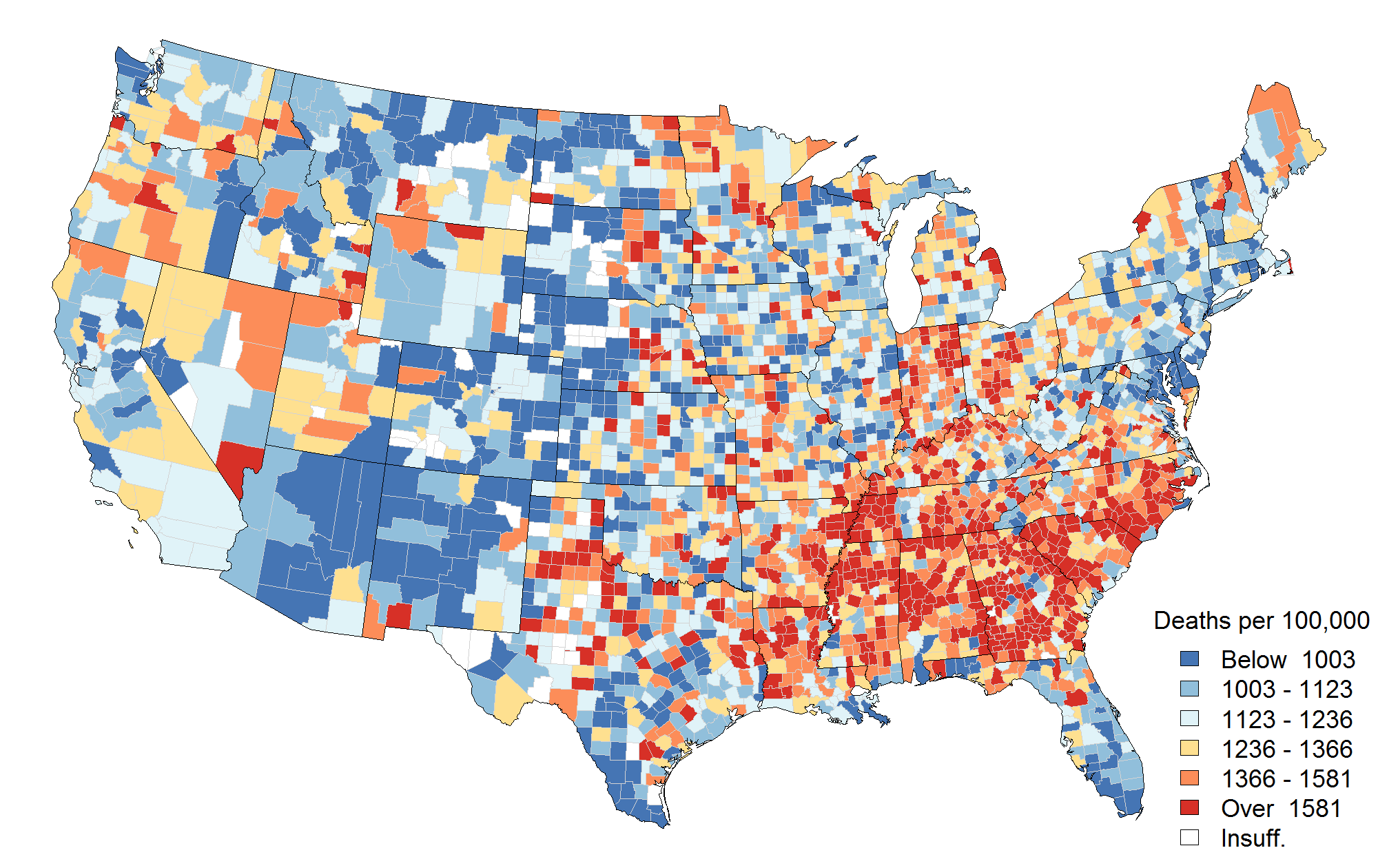}\label{fig:7a}}
        \subfigure[Ages 85+: 1973]{\includegraphics[width=.32\textwidth]{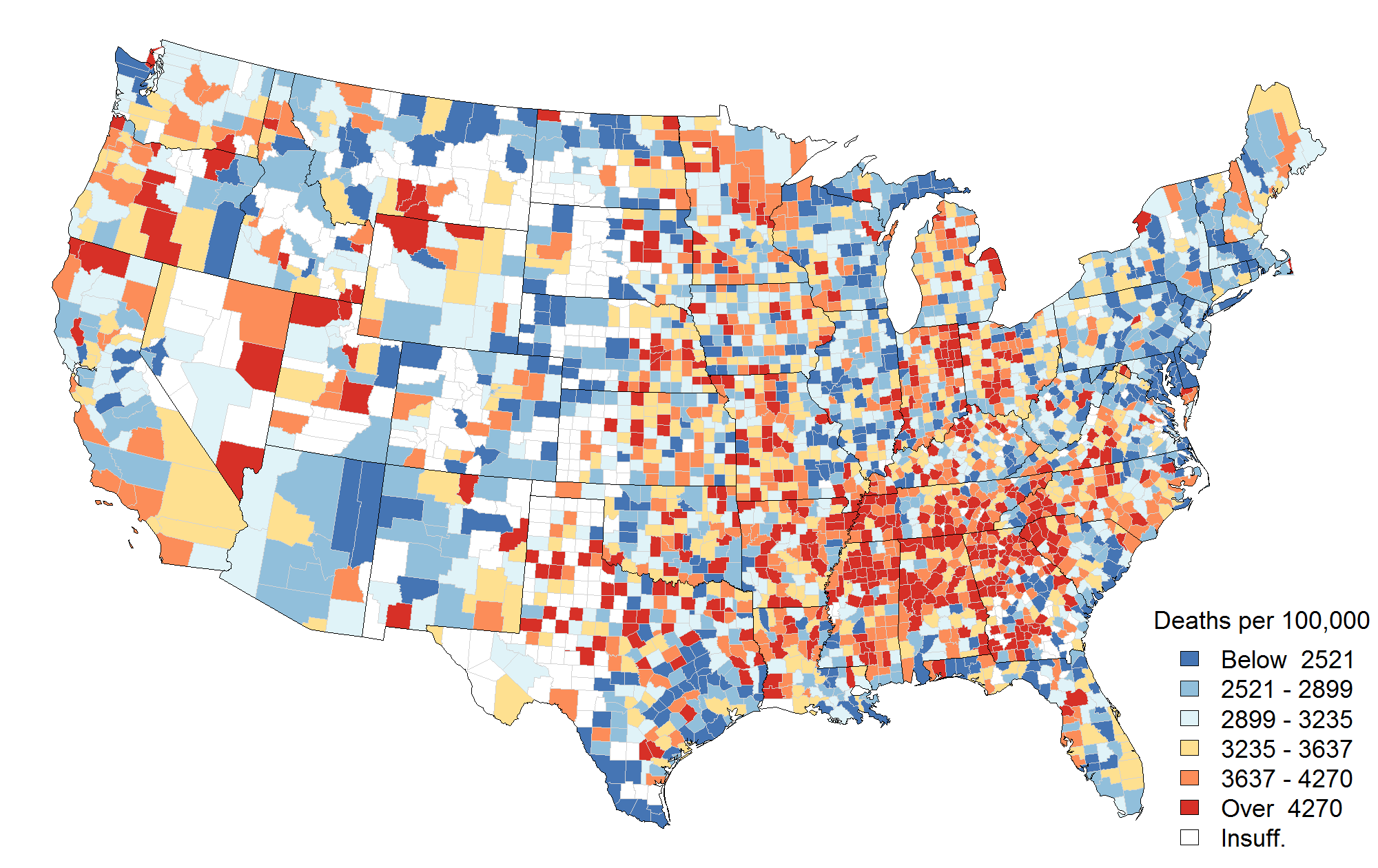}\label{fig:8a}}\\
        \subfigure[Ages 65--74: Declines]{\includegraphics[width=.32\textwidth]{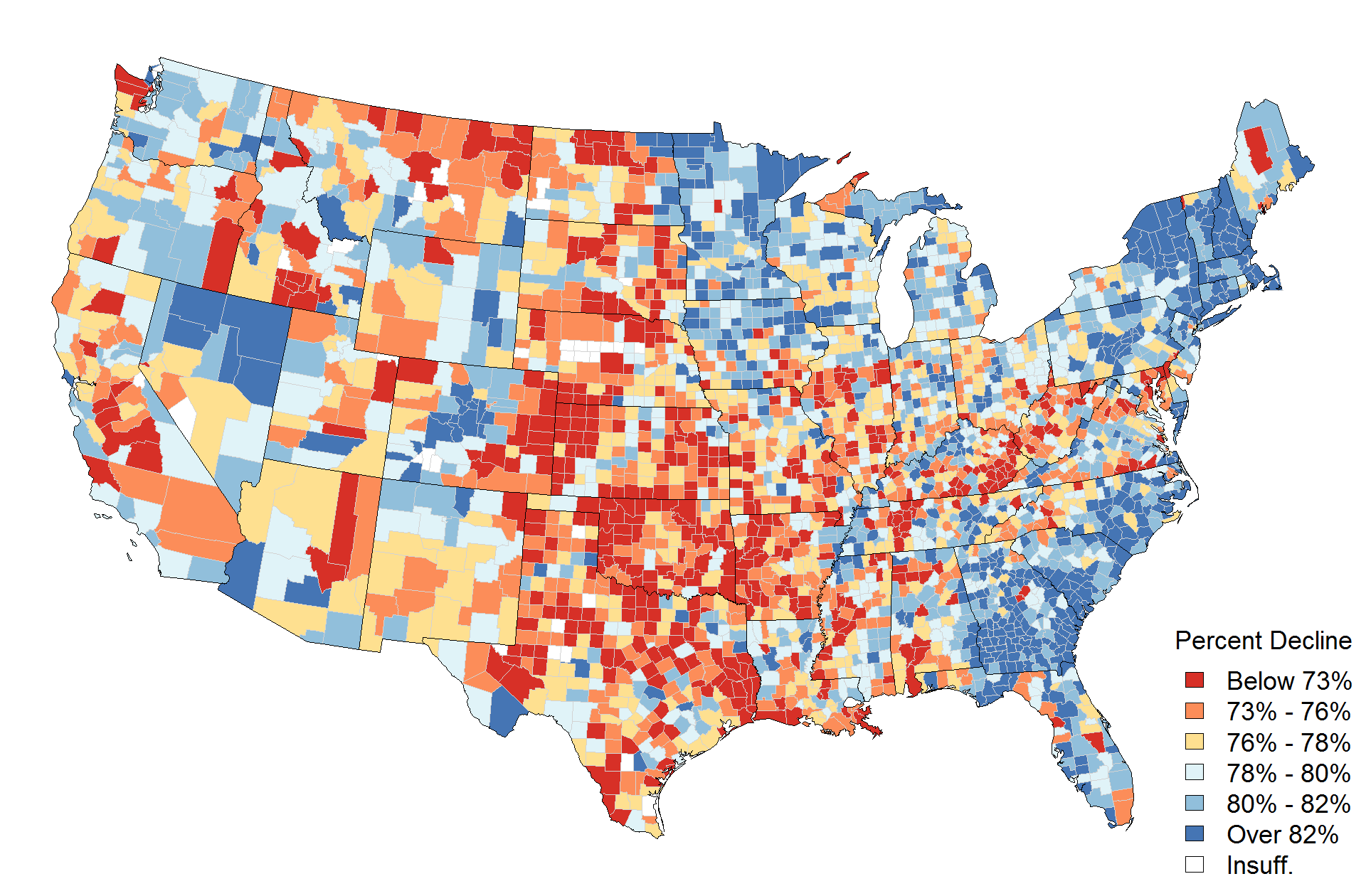}\label{fig:6d}}
        \subfigure[Ages 75--84: Declines]{\includegraphics[width=.32\textwidth]{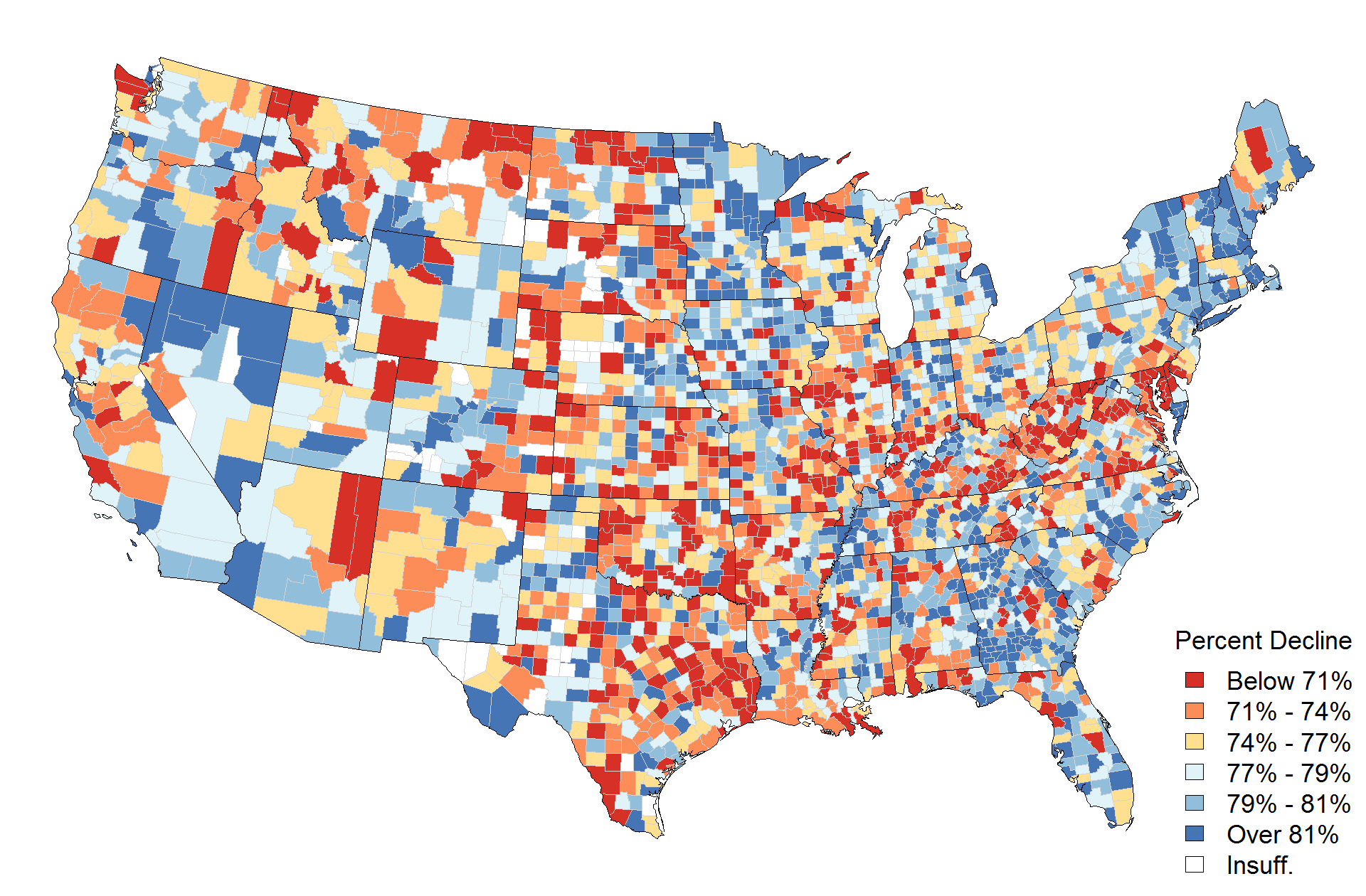}\label{fig:7d}}
        \subfigure[Ages 85+: Declines]{\includegraphics[width=.32\textwidth]{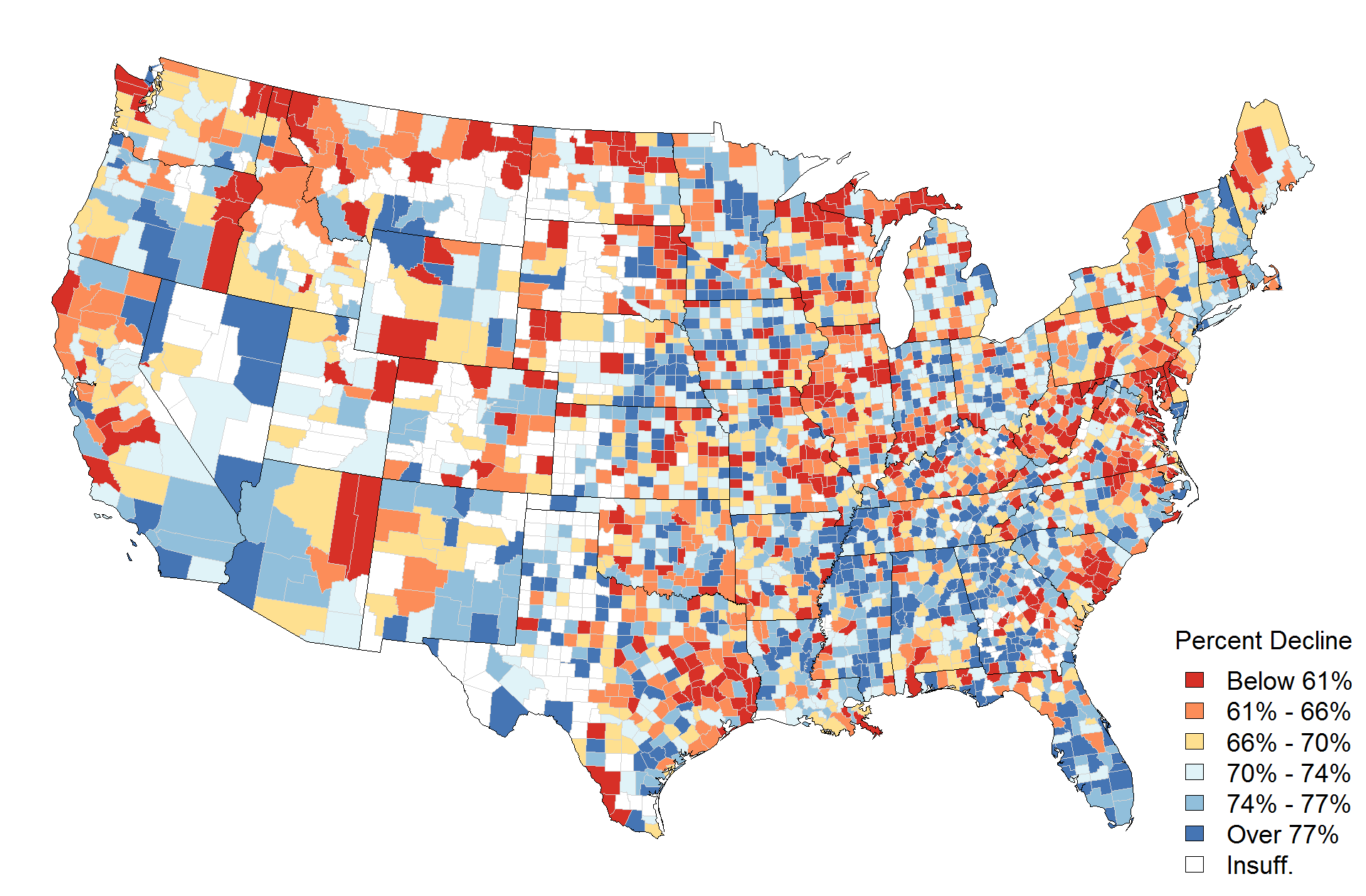}\label{fig:8d}}
    \end{center}
    \caption{Maps of the stroke mortality rates at the beginning of the study period (1973) and {the percent declines over the course of the study period}. Note that estimates for counties with fewer than 100 people in an age bracket in 1973 are suppressed.}
    \label{fig:rates}
\end{sidewaysfigure}

Turning our attention to the geographic patterns in stroke death rates presented in Figure~\ref{fig:rates}, we find substantial differences between age groups.
For the youngest subpopulation (ages 65--74), the clear geographic pattern shown in Figure~\ref{fig:6a} prominently highlights the so-called ``stroke belt'' in the rates from 1973, and the map of the percent declines in Figure~\ref{fig:6d} --- with large declines along the East Coast and smaller declines in the region stretching from Texas to the Dakotas --- seems to indicate the ``shift'' in the stroke belt identified by \citet{shiftingbelt}.
These patterns, however, are much less evident among the older age groups, especially for the eldest population.  Here, the rates in Figure~\ref{fig:8a} exhibit far less spatial clustering, while Figure~\ref{fig:8d} suggests that the rate of declines in mortality for those 85 and older was generally slower nationwide compared to those observed for those aged 65--74 and 75--84.

\section{Discussion}\label{sec:disc}
This paper has extended the $\MSTCAR$ model of \citet{quick:waller} to a generalized linear model for the purposes of analyzing a dataset comprised of county-level counts of stroke mortality, allowing for a nonseparable multivariate spatiotemporal dependence structure. Our analysis revealed spatiotemporal trends in stroke mortality that varied by age group, in addition to the nationwide reduction in 
rates previously noted in the literature \citep[e.g.,][]{gillum:2011:stroke,linda}.  We also observed differing aspects of the western shift in
parts of the South
for each age group, as 
identified in the total population by \citet{shiftingbelt}, and
explored the impact of non-linear trends in stroke mortality
via the SPY tool.

When modeling event rates for a rare event such as stroke mortality, it is important to leverage as much information as possible to achieve reliable estimates.
In addition to incorporating spatial structure into the model --- allowing for regional patterns such as the stroke belt to lend support to less populated counties --- the $\MSTCAR$ accounts for temporal correlation between observations in consecutive years and multivariate dependencies, such as those between observations in different age-brackets.
These additional sources of information can be invaluable when dealing with outlying counts, a problem which particularly plagues counties with small population sizes where an increase/decrease of a single death can dramatically change the observed mortality rate.
Overcoming this challenge will be paramount in our future work, where we wish to use the $\MSTCAR$ model to analyze
racial and gender disparities in heart disease and stroke mortality across various age groups.
While there is computational burden associated with implementing the $\MSTCAR$ for such large multivariate datasets, these analyses will provide incredible insight into these various disparities.

{Another area for future research is understanding the factors that contribute to differential geographic patterns by age group in both the baseline 1973 stroke mortality rates as well as the patterns of declining stroke mortality rates. 
While it is well known that the risk for stroke increases with age, the spatiotemporal patterns of stroke mortality by age group have not been documented previously.  Hypotheses for understanding the observed differential spatio-temporal patterns in stroke mortality by age group include, but are not limited to, the following categories: 1) spatio-temporal differences in the relative contributions of decreasing case fatality rates and incidence rates by age group {\citep[e.g.,][]{case_fatality,geo_variation}}; 2) differential influence of living conditions (e.g. socioeconomic resources, access to quality health care, access to healthy food and recreational environments, etc.) or changes in those living conditions, on stroke mortality by age group {\citep[e.g.,][]{tassone,neighborhood}}; or 3) differences in the accuracy of death certificate reporting by age group due to more co-morbidities and competing conditions of death in the older ages. 
}

Lastly, a topic worth discussing is that publicly available data for rare events such as deaths from stroke can be difficult to find due to data
privacy issues.  
While data such as the number of stroke-related deaths are publicly available at the county-level from NCHS (via CDC Wonder), subsets of data with less than 10 events in a geographic region are suppressed beginning in 1989 \citep{cdc:sharing}. 
This results in over {80}\% of the data points for those aged 65--74 and nearly 70\% of the over 380,000 data points used in this analysis being suppressed to the public.
To overcome this 
privacy issue while still preserving utility, \citet{quick:zero} have explored the 
risks associated with the generation of {synthetic} data 
for rare event data for a single population for a single year of data.
As an area of active research,
we hope to use the methodology proposed here to generate reliable synthetic public-use data for small areas 
which respects the spatial-, temporal-, and multivariate structures in the true data, thereby providing greater access to complete, high quality data.


\bibliographystyle{jasa}
\bibliography{cdc_ref,cdc_epi}

\begin{thebibliography}{37}
\newcommand{\enquote}[1]{``#1''}
\expandafter\ifx\csname natexlab\endcsname\relax\def\natexlab#1{#1}\fi

\bibitem[\protect\citename{Anderson et~al., }2001]{icd9:icd10}
Anderson, R.~N., Mini{\~n}o, A.~M., Hoyert, D.~L., and Rosenberg, H.~M. (2001).
\newblock \enquote{Comparability of cause of death between {ICD}--9 and
  {ICD}--10: Preliminary Estimates.}
\newblock {\em National Vital Statistics Reports\/}, 49.

\bibitem[\protect\citename{Banerjee et~al., }2014]{BCG}
Banerjee, S., Carlin, B.~P., and Gelfand, A.~E. (2014).
\newblock {\em Hierarchical Modeling and Analysis of Spatial Data\/}.
\newblock Boca Raton, FL: Chapman \& Hall/CRC.

\bibitem[\protect\citename{Bernardinelli et~al., }1995]{bernardinelli}
Bernardinelli, L., Clayton, D., and Montomoli, C. (1995).
\newblock \enquote{Bayesian estimates of disease maps: How important are
  priors?}
\newblock {\em Statistics in Medicine\/}, 14, 2411--2431.

\bibitem[\protect\citename{Besag, }1974]{besag}
Besag, J. (1974).
\newblock \enquote{Spatial interaction and the statistical analysis of lattice
  systems (with Discussion).}
\newblock {\em Journal of the Royal Statistical Society, Series B\/}, 36,
  192--236.

\bibitem[\protect\citename{Besag et~al., }1995]{besag:poisson}
Besag, J., Green, P., Higdon, D., and Mengersen, K. (1995).
\newblock \enquote{Bayesian computation and stochastic systems (with
  discussion).}
\newblock {\em Statistical Science\/}, 10, 3--66.

\bibitem[\protect\citename{Besag and Higdon, }1999]{besag:higdon}
Besag, J. and Higdon, D. (1999).
\newblock \enquote{Bayesian analysis of agricultural field experiments (with
  discussion).}
\newblock {\em Journal of the Royal Statistical Society: Series B\/}, 61,
  691--746.

\bibitem[\protect\citename{Besag and Kooperberg, }1995]{besag95}
Besag, J. and Kooperberg, C. (1995).
\newblock \enquote{On conditional and intrinsic autoregressions.}
\newblock {\em Biometrika\/}, 82, 733--746.

\bibitem[\protect\citename{Besag et~al., }1991]{bym}
Besag, J., York, J., and Molli\'e, A. (1991).
\newblock \enquote{Bayesian image restoration, with two applications in spatial
  statistics.}
\newblock {\em Annals of the Institute of Statistical Mathematics\/}, 43,
  1--59.

\bibitem[\protect\citename{Borhani, }1965]{strokebelt}
Borhani, N.~O. (1965).
\newblock \enquote{Changes and geographic distribution of mortality from
  cerebrovascular disease.}
\newblock {\em American Journal of Public Health\/}, 55, 673--681.

\bibitem[\protect\citename{Botella-Rocamora et~al., }2015]{b-r:2015}
Botella-Rocamora, P., Martinez-Beneito, M.~A., and Banerjee, S. (2015).
\newblock \enquote{A unifying modeling framework for highly multivariate
  disease mapping.}
\newblock {\em Statistics in Medicine\/}, 34, 1548--1559.

\bibitem[\protect\citename{Bradley et~al., }2014]{jon}
Bradley, J.~R., Wikle, C.~K., and Holan, S.~H. (2014).
\newblock \enquote{Mixed effects modeling for areal data that exhibit
  multivariate-spatio-temporal dependencies.}
\newblock ArXiv preprint, arXiv:1407.7479.

\bibitem[\protect\citename{Casper et~al., }1995]{shiftingbelt}
Casper, M., Anda, R.~F., Knowles, M., Pollard, R.~A., and Wing, S. (1995).
\newblock \enquote{The shifting stroke belt: Changes in the geographic pattern
  of stroke mortality in the {U}nited {S}tates.}
\newblock {\em Stroke\/}, 26, 755--760.

\bibitem[\protect\citename{CDC, }2003]{cdc:sharing}
CDC (2003).
\newblock \enquote{CDC/ATSDR Policy on Releasing and Sharing Data.}
\newblock Manual; Guide CDC-02. Available at
  \url{http://www.cdc.gov/maso/Policy/ReleasingData.pdf}.
\newblock Accessed June 30, 2015.

\bibitem[\protect\citename{Cressie and Huang, }1999]{cressie:huang}
Cressie, N. and Huang, H.-C. (1999).
\newblock \enquote{Classes of nonseparable, spatio-temporal stationary
  covariance functions.}
\newblock {\em Journal of the American Statistical Association\/}, 97,
  1330--1340.

\bibitem[\protect\citename{El-Saed et~al., }2006]{geo_variation}
El-Saed, A., Kuller, L.~H., Newman, A.~B., Lopez, O., Costantino, J., McTigue,
  K., Cushman, M., and Kronmal, R. (2006).
\newblock \enquote{Geographic variations in stroke incidence and mortality
  among older populations in four {US} communities.}
\newblock {\em Stroke\/}, 37, 1975--1979.

\bibitem[\protect\citename{Ergin et~al., }2004]{case_fatality}
Ergin, A., Muntner, P., Sherwin, R., and He, J. (2004).
\newblock \enquote{Secular trends in cardiovascular disease mortality,
  incidence, and case fatality rates in adults in the {U}nited {S}tates.}
\newblock {\em The American Journal of Medicine\/}, 117, 219--227.

\bibitem[\protect\citename{Gelfand and Vounatsou, }2003]{gelfand:mcar}
Gelfand, A.~E. and Vounatsou, P. (2003).
\newblock \enquote{Proper multivariate conditional autoregressive models for
  spatial data analysis.}
\newblock {\em Biostatistics\/}, 4, 11--25.

\bibitem[\protect\citename{Gillum et~al., }2011]{gillum:2011:stroke}
Gillum, R.~F., Kwagyan, J., and Obiesesan, T.~O. (2011).
\newblock \enquote{Ethnic and geographic variation in stroke mortality trends.}
\newblock {\em Stroke\/}, 42, 3294--3296.

\bibitem[\protect\citename{Gneiting, }2002]{gneiting2002}
Gneiting, T. (2002).
\newblock \enquote{Nonseparable, stationary covariance functions for space-time
  data.}
\newblock {\em Journal of the American Statistical Association\/}, 97,
  590--600.

\bibitem[\protect\citename{Howard et~al., }2001]{regards:declines}
Howard, G., Howard, V.~J., Katholi, C., Oli, M.~K., Huston, S., and Asplund, K.
  (2001).
\newblock \enquote{Decline in {US} stroke mortality: An analysis of temporal
  patterns by sex, race, and geogrpahic region.}
\newblock {\em Stroke\/}, 32, 2213--2218.

\bibitem[\protect\citename{Jin et~al., }2007]{JBC}
Jin, X., Banerjee, S., and Carlin, B.~P. (2007).
\newblock \enquote{Order-free co-regionalized areal data models with
  application to multiple-disease mapping.}
\newblock {\em Journal of the Royal Statistical Society, Series B\/}, 69,
  817--838.

\bibitem[\protect\citename{Klebba and Scott, }1980]{icd8:icd9}
Klebba, A.~J. and Scott, J.~H. (1980).
\newblock \enquote{Estimates of selected comparability ratios based on dual
  coding of 1976 death certificates by the eighth and ninth revisions of the
  {International Classification of Diseases}.}
\newblock {\em National Vital Statistics Reports\/}, 28.

\bibitem[\protect\citename{Knorr-Held, }2000]{knorr-held:2000}
Knorr-Held, L. (2000).
\newblock \enquote{Bayesian modelling of inseparable space-time variation in
  disease risk.}
\newblock {\em Statistics in Medicine\/}, 19, 2555--2567.

\bibitem[\protect\citename{Knorr-Held and Rue, }2002]{knorr-held:rue}
Knorr-Held, L. and Rue, H. (2002).
\newblock \enquote{On block updating in {M}arkov random field models for
  disease mapping.}
\newblock {\em Scandinavian Journal of Statistics\/}, 29, 597--614.

\bibitem[\protect\citename{Kochanek et~al., }2015]{deaths}
Kochanek, K.~D., Murphy, S.~L., and Xu, J. (2015).
\newblock \enquote{Deaths: Final data for 2011.}
\newblock {\em National Vital Statistics Reports\/}, 63, 3.

\bibitem[\protect\citename{Lisabeth et~al., }2006]{neighborhood}
Lisabeth, L.~D., {Diex Roux}, A.~V., Escobar, J.~D., Smith, M.~A., and
  Morgenstern, L.~B. (2006).
\newblock \enquote{Neighborhood environment and risk of ischemic stroke.}
\newblock {\em American Journal of Epidemiology\/}, 165, 279--287.

\bibitem[\protect\citename{Martinez-Beneito, }2013]{m-b:2013}
Martinez-Beneito, M.~A. (2013).
\newblock \enquote{A general modelling framework for multivariate disease
  mapping.}
\newblock {\em Biometrika\/}, 100, 539--553.

\bibitem[\protect\citename{NCHS, }2013]{census}
NCHS (2013).
\newblock \enquote{{Bridged-race population estimates: United States July 1st
  resident population by state, county, age, sex, bridged-race, and Hispanic
  origin}.}
\newblock \url{http://www.cdc.gov/NCHS/nvss/bridged_race.htm}.
\newblock Accessed: June 30, 2015.

\bibitem[\protect\citename{Quick et~al., }2013]{QBC}
Quick, H., Banerjee, S., and Carlin, B.~P. (2013).
\newblock \enquote{Modeling temporal gradients in regionally aggregated
  {C}alifornia asthma hospitalization data.}
\newblock {\em Annals of Applied Statistics\/}, 7, 154--176.

\bibitem[\protect\citename{Quick et~al., }2015{\natexlab{a}}]{hcar}
Quick, H., Carlin, B.~P., and Banerjee, S. (2015{\natexlab{a}}).
\newblock \enquote{Heteroscedastic conditional auto-regression models for
  areally referenced temporal processes for analysing {C}alifornia asthma
  hospitalization data.}
\newblock {\em Journal of the Royal Statistical Society, Series C\/}, 64,
  799--813.

\bibitem[\protect\citename{Quick et~al., }2015{\natexlab{b}}]{quick:zero}
Quick, H., Holan, S.~H., and Wikle, C.~K. (2015{\natexlab{b}}).
\newblock \enquote{Zeros and ones: A case for suppressing zeros in sensitive
  count data with an application to stroke mortality.}
\newblock {\em Stat\/}, 4, 227--234.

\bibitem[\protect\citename{Quick et~al., }2015{\natexlab{c}}]{quick:waller}
Quick, H., Waller, L.~A., and Casper, M. (2015{\natexlab{c}}).
\newblock \enquote{A nonseparable multivariate space-time model for analyzing
  county-level heart disease death rates for race and gender.}
\newblock ArXiv preprint, arXiv:1507.02741.

\bibitem[\protect\citename{Rue and Held, }2005]{rue:held}
Rue, H. and Held, L. (2005).
\newblock {\em Gaussian Markov Random Fields\/}.
\newblock Boca Raton, FL: Chapman \& Hall/CRC.

\bibitem[\protect\citename{Schieb et~al., }2013]{linda}
Schieb, L.~J., Mobley, L.~R., George, M., and Casper, M. (2013).
\newblock \enquote{Tracking stroke hospitalization clusters over time and
  associations with county-level socioeconomic and healthcare characteristics.}
\newblock {\em Stroke\/}, 44, 146--152.

\bibitem[\protect\citename{Stein, }2005]{stein2005}
Stein, M.~L. (2005).
\newblock \enquote{Space-time covariance functions.}
\newblock {\em Journal of the American Statistical Association\/}, 100,
  310--321.

\bibitem[\protect\citename{Tassone et~al., }2009]{tassone}
Tassone, E.~C., Waller, L.~A., and Casper, M.~L. (2009).
\newblock \enquote{Small-area racial disparity in stroke mortality.}
\newblock {\em Epidemiology\/}, 20, 234--241.

\bibitem[\protect\citename{Waller et~al., }1997]{waller:carlin}
Waller, L.~A., Carlin, B.~P., Xia, H., and Gelfand, A.~E. (1997).
\newblock \enquote{Hierarchical spatio-temporal mapping of disease rates.}
\newblock {\em Journal of the American Statistical Association\/}, 92,
  607--617.

\end{thebibliography}

\end{document}